# On-chip mechanical exceptional points based on an optomechanical zipper cavity

Ning Wu,[1,2,†] Kaiyu Cui,[1,2,*,†] Qiancheng Xu,[1,2] Xue Feng,[1,2] Fang Liu,[1,2] Wei Zhang,[1,2,3] and Yidong Huang[1,2,3,*]

[1]Department of Electronic Engineering, Tsinghua University, Beijing 100084, China

[2]Beijing National Research Center for Information Science and Technology (BNRist), Tsinghua University, Beijing 100084, China

[3]Beijing Academy of Quantum Information Science, Beijing, China

*Corresponding author: kaiyucui@tsinghua.edu.cn; yidonghuang@tsinghua.edu.cn

[†]These authors contributed equally to this work.

**Abstract:** Exceptional points (EPs) represent a distinct type of spectral singularity in non-Hermitian systems, and intriguing physics concepts have been studied with optical EPs recently. As a system beyond photonics, the mechanical oscillators coupling with many physical systems are expected to be further exploited EPs for mechanical sensing, topology energy transfer, nonreciprocal dynamics etc. In this study, we demonstrated on-chip mechanical EPs with a silicon optomechanical zipper cavity, wherein two near-degenerate mechanical breathing modes are coupled via a single co-localized optical mode. By tailoring the dissipative and coherent couplings between two mechanical oscillators, the spectral splitting with 1/2 order response, a distinctive feature of EP, was observed successfully. Our work provides an integrated platform for investigating the physics related to mechanical EPs on silicon chips and suggests their possible applications for ultrasensitive measurements.

# 1. Introduction

The non-Hermitian system, which exchanges energy with the outside environment, is quite different from the conservative Hermitian system. Exceptional points (EPs) are special degenerate points of spectra that exist in the non-Hermitian system. For $N$th-order EPs, $N$ eigenvalues and eigenvectors coalesce simultaneously at the EPs, and the theorem of completeness and orthogonality fails in this system[1,2,3]. Thus, owing to this characteristic, intriguing physics concepts are expected. In the past few years, optical EPs have been demonstrated in various platforms including optical and microwave cavities[4,5,6,7,8], photonic crystal slabs[9], and multilayered plasmonic structures[10]. Furthermore, certain counterintuitive phenomena have been reported and observed experimentally[1,2,3]. For instance, when a small disturbance of strength $\varepsilon$ acts on the EPs, spectral splitting is proportional to $\varepsilon^{1/N}$. As a result, for a $1/N$ order response, the spectral splitting near the EPs may be far greater than the normal mode splitting in the Hermitian system, where splitting of degenerate points is proportional to the perturbation $\varepsilon$. This property was utilized to improve the performance of the single-mode operation in multimode laser cavities[11] and enhance the sensitivities of sensors[4,10,12,13,14]. In addition, encircling the EPs in parameter space is nonreciprocal and the counter-clockwise evolution is distinct from the clockwise one. This effect was demonstrated in topology energy transfer[15,16] and asymmetric mode switching[17]. However, studies to date have focused primarily on optical EPs. The EPs in other physical systems are expected to be exploited further.

In cavity optomechanics, mechanical properties such as resonant frequency and dissipation rate can be adjusted effectively via optical modes through radiation pressure. Moreover, the mechanical and optical modes coupled with the thermal bath constitute a natural non-Hermitian system. Although this system is promising for manipulating mechanical modes and achieving mechanical EPs in multi-physics systems, experimental studies on mechanical EPs are still very limited[15,16,18]. Fully integrated mechanical EPs have remained elusive because of great challenges in on-chip mechanical manipulation.

In this study, we demonstrate on-chip mechanical EPs in an optomechanical zipper cavity at ambient environment, wherein two near-degenerate GHz mechanical breathing modes are coupled via a co-localized single optical mode. In this experiment, the strength of dissipative and coherent couplings between two mechanical oscillators are controlled by adjusting the frequency and power of the pump light to compensate for the difference of complex frequencies between the two mechanical breathing modes. Accordingly, the topology surface near the mechanical EP was mapped and a spectral splitting with 1/2 order response, a distinctive feature of EPs, was observed. As nanomechanical resonators can be coupled with many physical systems[19] and are suitable for detecting quantities such as mass, charge, and torque[20,21,22], our work paves the way for high-sensitivity measurement with mechanical EPs on integrated platforms. Moreover, this research sets the foundation for studying the related physics of mechanical EPs and other non-Hermitian phenomena based on optomechanical crystals.

## 2. Results

Mechanical EPs based on multimode optomechanical coupling can be analyzed as two independent pendulums with different oscillation frequencies (two mechanical modes) interacting with an optical mode in a Fabry-Perot cavity constructed by the two pendulums, as shown schematically in Fig. 1a. The optical mode acts as a bridge that connects the two mechanical modes. The oscillation of each pendulum can modulate the resonant frequency of the optical mode, thus subsequently, the light intensity in the cavity changes. Accordingly, the oscillation of radiation pressure force that originates from the variation of light intensity in the cavity affects the motion of the other pendulum and thereby enables the coupling between the two mechanical modes. It should be noted that the coupling here is distinct from mode coupling in the Hermitian system. In a Hermitian system, one of the mechanical oscillators exerts a force proportional to its displacement on another oscillator and leads to the coherent coupling[23,24,25]. When enhancing the coherent coupling rate, the normal mode splitting increases. However, for the non-Hermitian optomechanical system, in addition to the coherent coupling, the force exerted by one of the mechanical oscillators on the other also contains the dissipative coupling,[26] which results in a significant change in the physical scenario. The corresponding quantitative description is as follows.

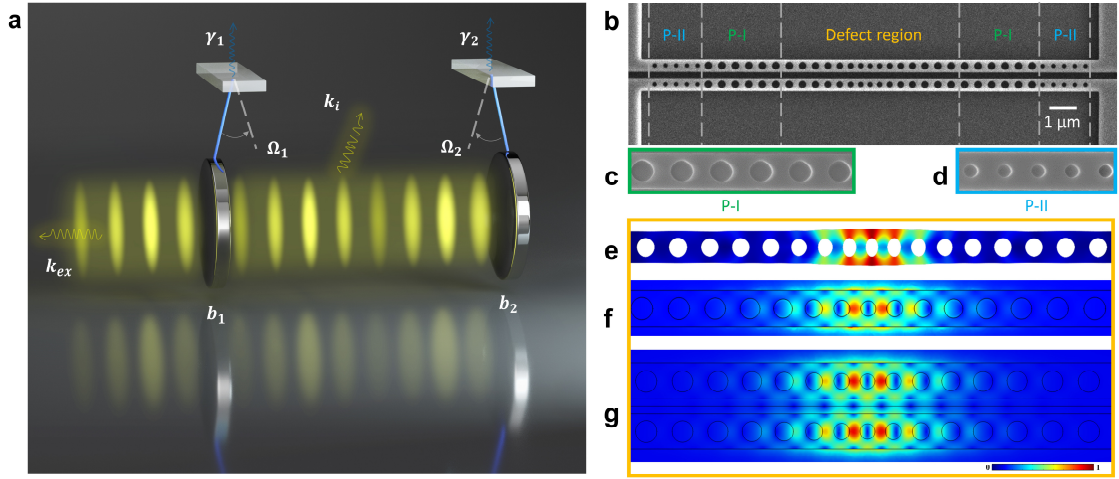

**Figure 1 a,** Schematic of a Fabry-Perot cavity consists of two pendulums as the reflecting mirror that act as mechanical modes $b_1$ and $b_2$ with oscillation frequency, $\Omega_1$ and $\Omega_2$, and are coupled to the thermal bath at the rate of $\gamma_1$ and $\gamma_2$, respectively. The optical mode of the Fabry-Perot cavity loses energy via intrinsic loss channel at rate $k_i$ and detectable extrinsic coupling channel at rate $k_{ex}$. **b,** Scanning electron microscope (SEM) image of silicon optomechanical zipper cavity. Magnified SEM image of **(c)** periodic structure in P-I region and **(d)** periodic structure in P-II region. **e,** Displacement field of the breathing mode simulated with the finite element method (FEM) in one arm of the optomechanical zipper cavity. FEM simulation of electric field |E| of **(f)** first order optical mode in a single nanobeam cavity and **(g)** first order odd optical mode in zipper cavity.

The coupling strength between the mechanical oscillators, including both the coherent and dissipative couplings, is primarily controlled by the optical mode. When a pump laser with frequency $\omega_L$ drives the optical mode with resonance frequency $\omega_{cav}$, detuning $\Delta=\omega_L-\omega_{cav}$ and intracavity photon number $n_{cav}$ can be adjusted by changing frequency $\omega_L$ and power $P_{in}$ of the pump laser, respectively. To further quantitatively describe the relationship between these parameters and the dynamics of the model shown in Fig. 1a, the Heisenberg-Langevin equations[27] are solved in the frequency domain (see Supplementary Section I), and the effective Hamiltonian of the mechanical oscillators can be expressed as

$$H_{eff} = \begin{bmatrix} \Omega_1 - i\gamma_1/2 + g_1^2\chi & g_1 g_2 \chi \\ g_1 g_2 \chi & \Omega_2 - i\gamma_2/2 + g_2^2\chi \end{bmatrix} \quad (1)$$

where $\chi \approx n_{cav}(1/((w+\Delta)+ik/2)-1/((w-\Delta)+ik/2))$ and $w=(\Omega_1+\Omega_2)/2$. $k$ is the optical dissipation rate and $g_j$ ($j$=1, 2) is the vacuum optomechanical coupling rate between the optical mode and the $j$th mechanical mode. For the diagonal elements of the effective Hamiltonian $H_{eff}$, mechanical resonant frequencies $\Omega_1$ and $\Omega_2$ are modified owing to the optical spring effect. Meanwhile, the mechanical dissipation rates $\gamma_1$ and $\gamma_2$ are decreased (increased) by the anti-damping (damping) effect of optomechanical coupling when a blue (red) detuned laser[27,28] is used. Further, the off-diagonal element is a complex number, and its real and imaginary parts correspond to the coherent and dissipative couplings, respectively. The eigenvalues, which characterize the evolution of this two-level system, can be further represented as

$$\tilde{\omega}_{\pm} = \omega_{\pm} - i\gamma_{\pm}/2 = \frac{1}{2}(\Omega_1 - i\gamma_1/2 + \Omega_2 - i\gamma_2/2) + \frac{1}{2}(g_1^2 + g_2^2)\chi \pm \frac{1}{2}S \tag{2}$$

where $S = \sqrt{\left((\Omega_1 - i\gamma_1/2 + g_1^2\chi) - (\Omega_2 - i\gamma_2/2 + g_2^2\chi)\right)^2 + 4(g_1 g_2 \chi)^2}$

Herein, the resonant frequencies and mechanical dissipation rates (linewidth) of both mechanical modes are modified by $\frac{1}{2}(g_1^2 + g_2^2)\chi$, and the mode splitting between them is controlled and reflected by parameter $S$. By tuning parameter $\chi$ via sweeping of the detuning $\Delta$ and intracavity photon number $n_{cav}$, the eigenvalue and eigenvector coalesce at the EPs where $S=0$ when parameter $\chi$ satisfies the following condition

$$\chi_{eps}(\Delta_0, n_{cav_0}) = \frac{(\Omega_1 - \Omega_2) - i\frac{\gamma_1 - \gamma_2}{2}}{g_2^2 - g_1^2 \pm i 2 g_1 g_2} \tag{3}$$

Typically, parameter $\chi_{eps}$ is a complex number at the EPs, and indicates that the coherent and dissipative coupling mechanisms are both necessary. It can be seen that $\chi_{eps}$ is an imaginary number when $g_1=g_2$ and $\gamma_1=\gamma_2$, which indicates that the dissipative coupling alone is sufficient to realize the EPs related to anti-parity time symmetry[29,30,31] in this situation. Although the hybridization of two mechanical modes was reported recently in the unresolved-sideband regime ($k>>\Omega$)[23,24,25], the dissipative coupling is strongly diminishes comparing with the coherent coupling in this regime[27]. As a result, it is crucial to investigate the resolved-sideband regime ($k<<\Omega$) of the EPs wherein the coherent and dissipative couplings can be adjusted independently and effectively to compensate for the difference of resonant frequencies and mechanical dissipation rates simultaneously.

When certain disturbances act on a system at the EPs, parameter $\chi$ may deviate from $\chi_{eps}$. We define perturbation strength $\varepsilon$ as $\varepsilon=\chi-\chi_{eps}$ to characterize the deviation from the EPs. The complex frequency splitting $S$ can be further represented as a function of perturbation $\varepsilon$, $S = \sqrt{|A\varepsilon + B\varepsilon^2|}e^{i\varphi}$, where $A$ and $B$ are complex constants and parameter $\varphi$ reflects the ratio of splitting on the real and imaginary parts. For a small disturbance ($B\varepsilon^2 \ll A\varepsilon$), $|S| \approx \sqrt{|A\varepsilon|}$. In particular, the amplitude of the coefficient $A$ will scale with $\Omega_1-\Omega_2$ if $\gamma_1=\gamma_2$. This implies that a small perturbation $\varepsilon$ results in large mode splitting at EPs owing to the 1/2 order response and the large difference of mechanical resonant frequencies.

Following the multimode optomechanical coupling model introduced above, we propose and design the optomechanical zipper cavity, consisting of two identical silicon nanobeam cavities with a 200 nm gap between them, as shown in Fig. 1b. As one arm of the optomechanical zipper cavity, a nanobeam cavity consists of three parts: two quasi-periodic mirror regions (P-I and P-II) and one defect cavity region. The first quasi-periodic structure P-I (Fig. 1c) can avoid optical energy loss via a waveguide as the optical resonant frequency is on the center of the photonic band gap. The second quasi-periodic structure P-II (Fig. 1d) is used to regulate the mechanical radiation loss when the phononic bandgap is designed properly. In the defect region, the radius of holes increases gradually from the center moving outward, and both the optical and mechanical modes can be confined[32,33]. We focus on the mechanical breathing mode (Fig. 1e) and the first-order optical mode (Fig. 1f). When two identical nanobeam cavities approach each other to constitute the zipper cavity, the mechanical breathing mode in each nanobeam cavity remains independent of each other. In contrast, the first-order optical degenerate modes in each nanobeam cavity couple to each other through the evanescent field and result in renormalized odd (Fig.1g) and even optical modes with distinct resonant frequencies. In principle, either of these optical modes can be used to realize the mechanical EPs when it couples to mechanical breathing modes via radiation pressure. In our design, a low optical dissipation rate is required because the system should reach the resolved-sideband regime. Moreover, both high vacuum optomechanical coupling rate $g_j$ and low optical dissipation rate $k$ are expected to decrease the intracavity photon number $n_{cav}$ at EPs and diminish the influence of other nonlinear effects related to the light absorption[34,35]. Considering these objectives, the structure parameters such as the radius and pitches of holes were optimized. Finally, the designed optical wavelength was $\lambda_{odd}$=1542.5 nm ($\lambda_{even}$=1551 nm), and the optical linewidth was $k/2\pi$=0.04 GHz (0.12 GHz) corresponding to an optical quality factor $Q_o$=4.86×10$^6$ (1.6×10$^6$). In addition, the designed mechanical frequency was $\Omega_1/2\pi=\Omega_2/2\pi$=5.634 GHz and the optomechanical coupling rate was $g_1/2\pi=g_2/2\pi$=0.63 MHz.

After determining the parameters of the zipper cavity, the designed cavity was fabricated on a silicon-on-insulator wafer having top layer thickness of 220 nm and a 3-μm thick buried-oxide layer (see Supplementary Section III). Figure 2a shows the measurement setup used to characterize the evolution of mechanical spectra with tuned optical parameters. A tapered fiber controlled with nanopositioners was used to evanescently couple the input blue-detuned laser into the zipper cavity. In this experiment, a pump-probe scheme[36,37,38] was used to monitor the detuning Δ and the optical dissipation rate $k$ (see Supplementary Section IV).

The optical and mechanical spectra are illustrated in Fig. 2b-d. In Fig. 2b, the first resonant dip at 1534.9 nm is an odd optical mode while the second resonance dip at 1543 nm corresponds to an even optical mode. In this experiment, we excite the optical odd mode to realize the mechanical EPs, and its Lorentzian fitting is shown in the inset. The corresponding optical linewidth of $k/2\pi=1$ GHz ($Q_o=1.95\times10^5$) indicates that the resolved-sideband regime ($\Omega \gg k$) was reached. The derived extrinsic coupling rate of $k_{ex}/2\pi= 0.208$ GHz from Fig. 2b is assumed to be constant during the experiment and was used to calculate the intracavity photon number $n_{cav}$. The normalized amplitude response of $S_{12}$ (red points in Fig. 2c) detected via the vector network analyzer (VNA) after calibrating the frequency response (see Supplementary Section V) is consistent with the theoretical fitting (black line), and the optical linewidth and detuning can be derived. Meanwhile, a narrow resonant dip appears when the beat frequency between the pump and weak probe lights is close to the mechanical resonant frequency. This can be attributed to the effect of optomechanical induced absorption (OMIA)[38]. Here, with the large $n_{cav}$ inside the cavity, the derived optical linewidth $k/2\pi= 1.26$ GHz is distinct from the result in Fig. 2b because of the nonlinear absorption effect[33]. The power spectrum density (PSD) in Fig 2d carrying the information of mechanical motion and the related mechanical mode properties can be deduced by using multiple spectra measured under different detuning and optical input power. Here, the fitting parameters were as follows $\Omega_1/2\pi=5.6531$ GHz, $\Omega_2/2\pi=5.6581$ GHz, $Q_1=\Omega_1/\gamma_1=1100$, $Q_2=\Omega_2/\gamma_2=1440$, $g_1=0.72$ MHz, and $g_2=0.32$ MHz. Compared to the intrinsic mechanical resonant frequencies, resonant peaks in Fig 2d are blue shifted due to the optical spring effect. Further, the mechanical peak at higher frequency is weaker than that at lower frequency because the optomechanical coupling rate $g_2$ are lower. In addition, the discrepancy of the intrinsic resonant frequencies between the two mechanical oscillators originates from the fabrication error and it contributes to a large eigenvalue splitting near the EPs.

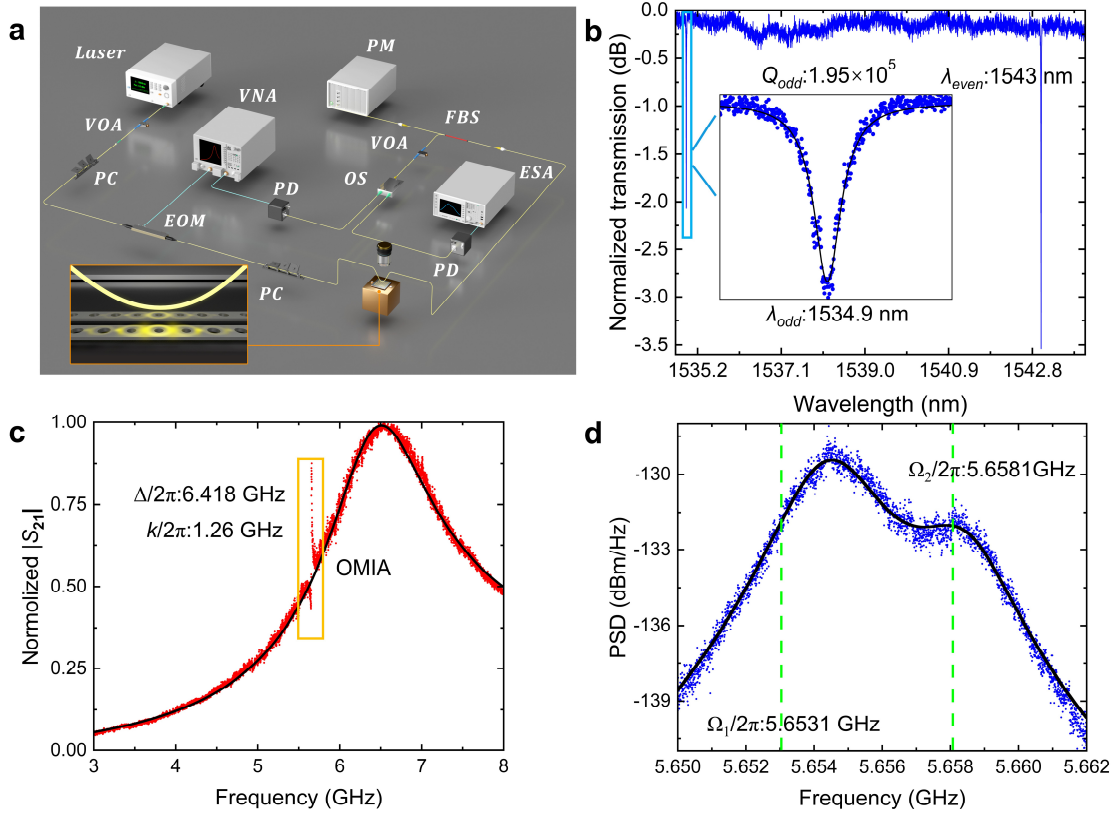

**Figure 2 a,** Experimental setup schematic. VOA: variable optical attenuator; PC: polarization controller; EOM: electro-optic modulator; VNA: vector network analyzer; FBS: fiber beam splitter; PM: power meter; OS: optical switch; PD: photodetector; ESA: electric spectrum analyzer. **b,** Low input power optical transmission spectrum. First (1534.9 nm) and second (1543 nm) resonant dips correspond to odd and even optical modes, respectively. **c,** Amplitude response of $S_{21}$. $k$ at high optical power and detuning $\Delta$ are deduced from this response. **d,** Power spectrum density (PSD) of mechanical spectrum after subtracting the background noise, obtained at the same condition in (**c**). Green dashed-dot line represents intrinsic resonant frequency of both mechanical oscillators. (b)-(d) Black solid lines represent fitting results.

To analyze the evolution of the eigenvalue of the mechanical modes with the variation of the detuning $\Delta$ and the intracavity photon number $n_{cav}$, the mechanical spectra are measured by scanning the optical detuning $\Delta$ at a fixed laser power $P_{in}$=11 dBm and $P_{in}$=13 dBm in Fig 3a and Fig 3b, respectively. For $P_{in}$=11 dBm in Fig 3a, we observed a significant increase of the peak of the mechanical mode with lower frequency as the detuning $\Delta$ decreases, which indicates the decrease of the mechanical linewidth. Subsequently, the phonon lasing occurred in this lower frequency mode. In contrast, for $P_{in}$=13 dBm, it can be found that the peak (linewidth) of the mechanical mode with higher frequency shown in Fig 3b increases (decreases) dramatically as the detuning $\Delta$ decreases. Thereafter, the phonon lasing occurred in this higher frequency mode. This phenomenon is also an evidence of the coupling between the two mechanical oscillators. Otherwise, the lower frequency mode would always enter the phonon lasing regime first due to its higher optomechanical coupling rate $g_1$.

Furthermore, mechanical eigenvalues $\omega_{\pm} - i\gamma_{\pm}/2$ can be obtained from the mechanical spectra. Figure 3c and 3e (3d and 3f) display the evolution of the real and imaginary part of eigenvalues for different detuning $\Delta$ using the spectra in Fig. 3a (Fig. 3b), respectively. Here the experimental data (square markers) are consistent with the theoretical curves (solid lines) with the linear approximation. Additionally, the crossing (anti-crossing) of the dissipation rate $\gamma_{\pm}/2\pi$ and the anti-crossing (crossing) of frequency $\omega_{\pm}/2\pi$ of the mechanical modes appear as theoretical prediction when sweeping the detuning at $P_{in}$=11 (13) dBm. Nevertheless, the system will enter the phonon lasing regime (see Supplementary Section VIII) before the frequency crossing at $P_{in}$=13 dBm experimentally, owing to the low intrinsic mechanical dissipation rate of this test sample.

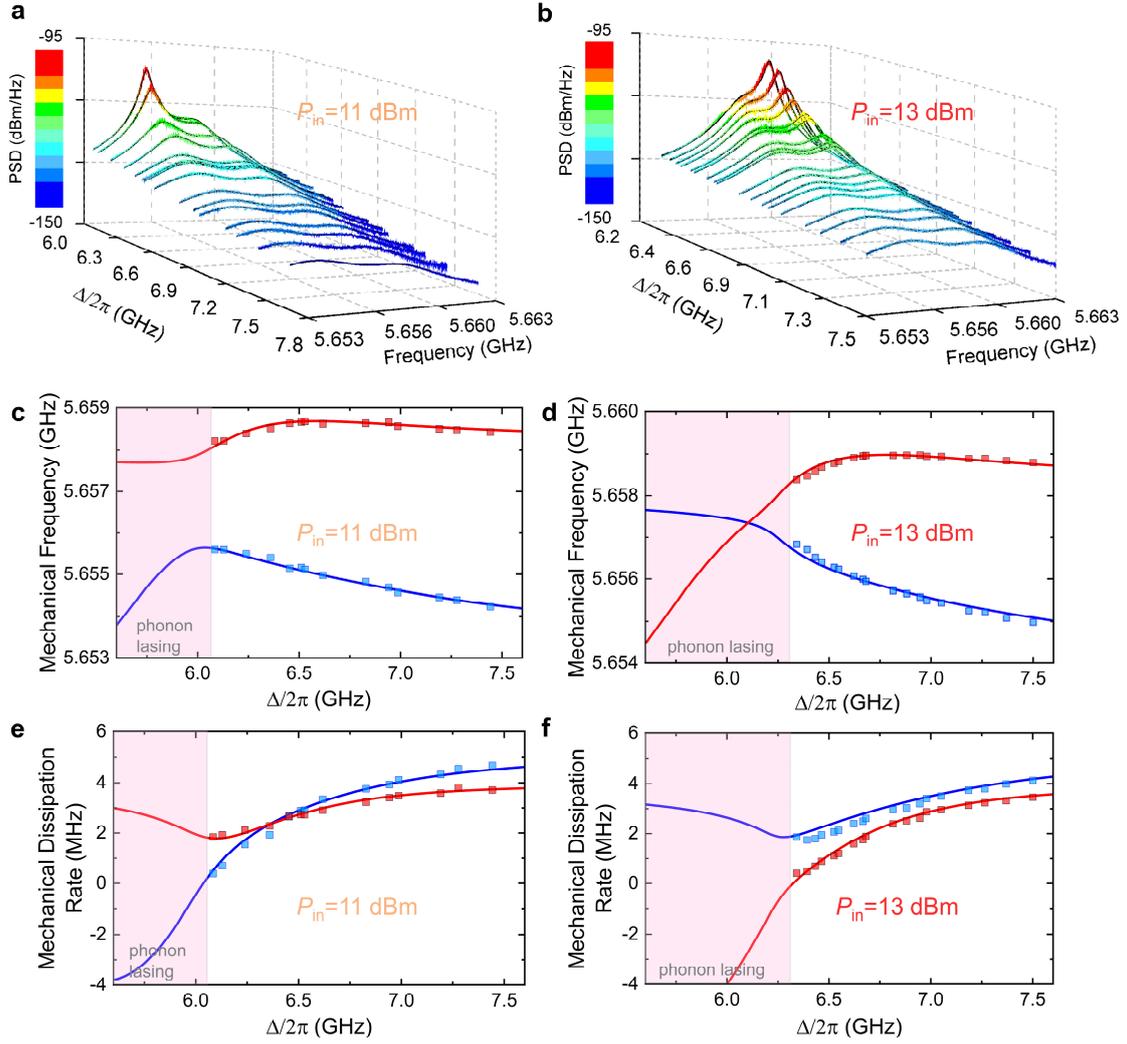

**Figure 3** (**a**) and (**b**) Power spectrum density (PSD) of mechanical spectra after subtracting the background noise when scanning optical detuning with fixed laser power $P_{in}$=11 dBm and $P_{in}$=13dBm, respectively. Black solid lines represent fit to experimental data. (**c**) and (**d**) Mechanical resonant frequencies $\omega_{\pm}/2\pi$ versus the optical detuning $\Delta$, deduced from the spectra in (a) and (b), respectively. (**e**) and (**f**) Mechanical dissipation rate $\gamma_{\pm}/2\pi$ versus the optical detuning $\Delta$, deduced from the spectra in (a) and (b), respectively; (c)-(f) square markers correspond to fitting results of experimental spectra, blue and red lines are theoretical results.

Following the above-mentioned procedure, we further reconstruct the topological surface of the mechanical eigenvalues under different laser powers and detuning. The mechanical resonant frequency $\omega_{\pm}/2\pi$ and the mechanical dissipation rate $\gamma_{\pm}/2\pi$ versus the detuning $\Delta$ and intracavity photon number $n_{cav}$, are plotted in Fig. 4a-b, respectively. It can be seen that the experimental data (blue and red points) are consistent with the theoretical surface results and the region near the EPs was reached. Here the relation between optical dissipation rate $k$ and intracavity photon number $n_{cav}$ has been considered in the theoretical calculation (see Supplementary Section VI), and the deviation between the experimental data and the theoretical surface mainly comes from the long-term drift of the intrinsic mechanical frequency (see Supplementary Section VII).

Figure 4c displays the distribution of the experimental data from Fig. 4a-b in the parameter space with the circle markers. Here the black solid line corresponds to parameters where the real parts of the eigenvalues are degenerate in Fig. 4a, while the black dashed line corresponds to the parameters where the imaginary parts of the eigenvalues are degenerate in Fig. 4b. As the real and imaginary parts of the mechanical modes are simultaneously degenerate at the EPs, the corresponding parameter point of the EPs is the intersection of the black solid and dash line. Due to this property of the EPs, at low laser power, the evolution of parameter will always cross the black dash line when sweeping the detuning. Then, the crossing of the dissipation rate of mechanical modes will show up. Similarly, when the input optical power is higher than the power required to reach the EPs, the evolution of parameters will cross the black solid line when sweeping the detuning, so that the crossing of mechanical frequency appeared and the phase change occurs at the EPs. The results in Fig. 3c-d and Fig. 3e-f are in the corresponding two regions, respectively.

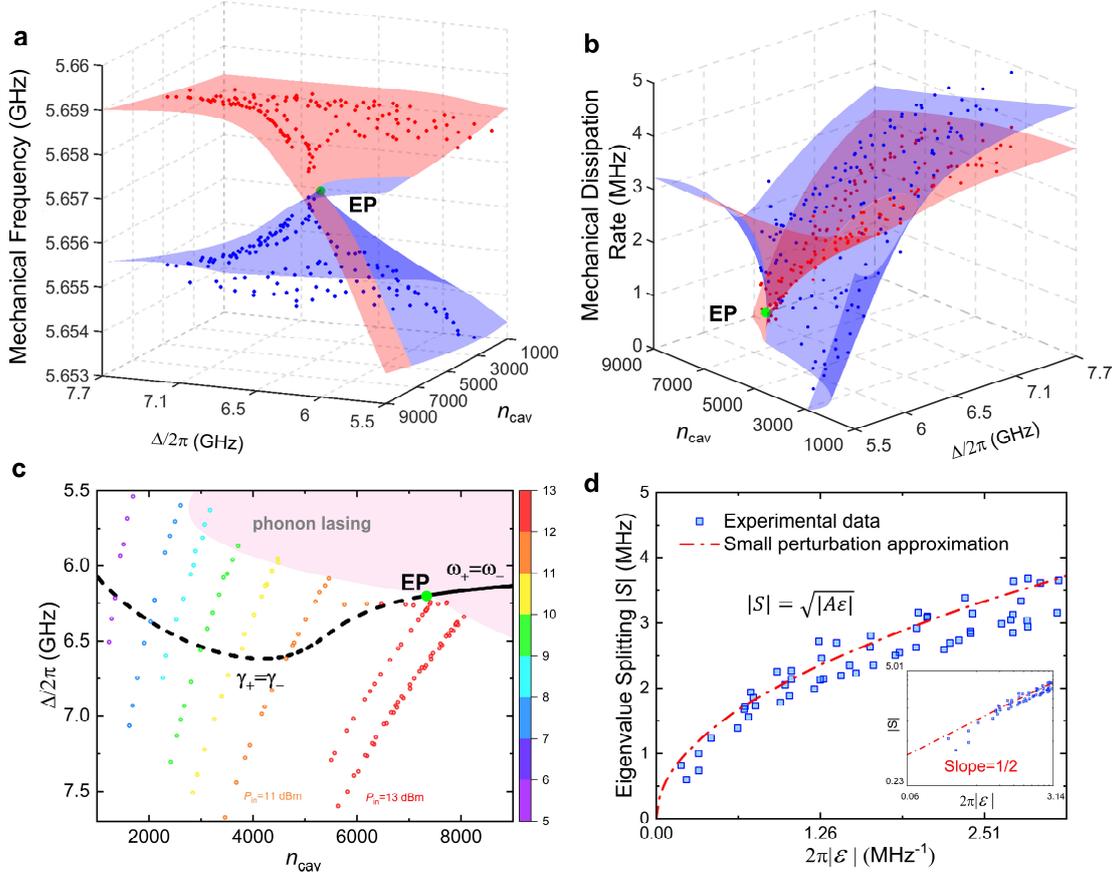

**Figure 4 (a)** Resonant frequencies $\omega_\pm/2\pi$ and **(b)** Dissipation rate $\gamma_\pm/2\pi$ of mechanical modes versus detuning $\Delta$ and intracavity photon number $n_{cav}$, respectively. In (a)-(b), topology surfaces are calculated from theoretical model using parameters deduced from experimental results. Blue and red points are experimental results. **(c)** the distribution of the experimental data in parameter space. **(d)** Amplitude of eigenvalue splitting $S$ versus amplitude of perturbation $\varepsilon$. The inset shows alternative logarithmic scale presentation where the 1/2 order response corresponds to the line with slope=1/2. (a)-(c) Green point corresponds to an EP.

Then, we focus on the variation of eigenvalues near the EPs. According to the theoretical model, the amplitude of mode splitting $S$ is a nonlinear function of the perturbation $\varepsilon$, where $|S|=\sqrt{|A\varepsilon+B\varepsilon^2|}$. As the experimental data are close to the EPs in the parameter space, the perturbation $\varepsilon$ of can be very small. This leads to a good agreement between the experimental data and the approximate theoretical response of $|S|=\sqrt{|A\varepsilon|}$ in Fig. 4d. Consequently, the 1/2 order response of the EPs was observed in this test sample. In addition, as the amplitude of perturbation $\varepsilon$ is a nonlinear function of detuning $\Delta$, the 1/2 order response and the nonlinear transduction between perturbation $\varepsilon$ and detuning $\Delta$ resulted in a sudden change of mechanical frequency splitting near the EPs in Fig. 4a when adjusting the detuning $\Delta$.

## 3. Discussion

In our experiment, a larger difference of mechanical frequency $|\Omega_1-\Omega_2|$ will further enhance the frequency splitting near the EPs even though the mechanical dissipation rate needs to be increased to avoid the effect of phonon lasing[39,40,41]. In addition, the device also benefits from the reduced mechanical linewidth by using a blue-detuned laser. What's more, owing to the self-reference detection scheme of frequency splitting, this system is robust to the mechanical frequency drifts[42]. In contrast, the sensor proposed by a single optomechanical coupling suffers from frequency drift[43,44]. Therefore, the device is suitable for realizing high-sensitivity sensors by combining the anti-damping modes and the mechanical EPs. It is noted that this work also offers a reliable, integratable platform for studying and utilizing multimode non-Hermitian physics. For instance, further using the red detuning laser might be suitable for the investigation of multimode optomechanical cooling[45] and the nonreciprocal dynamics of mechanical EPs[16,18] under the quantum ground state as the high frequency of mechanical breathing modes result in low thermal phonons; In addition, high order EPs[46] and the multimode dynamics in the phonon lasing regime[47] can also be explored further based on this platform.

**Funding**

This research was funded by the National Key R&D Program of China (Contract No. 2018YFB2200402); National Natural Science Foundation of China (Grant No. 91750206, 61775115); Beijing Municipal Science Technology Commission Z201100004020010; Beijing National Science Foundation contract Z180012; Beijing Frontier Science Center for Quantum Information; and Beijing Academy of Quantum Information Sciences.

**Acknowledgments**

The authors express their gratitude to Tianjin H-Chip Technology Group Corporation, Innovation Center of Advanced Optoelectronic Chip and Institute for Electronics and Information Technology in Tianjin, Tsinghua University for their fabrication support with electron beam lithography (EBL) and inductively coupled plasma (ICP) etching.

**Disclosures**

The authors declare no conflicts of interest.

**Data availability**

The data that support the plots within this paper and other findings are available from the corresponding author upon reasonable request.

**Author contributions**

N.W., K.C. contributed equally to this work. K.C. and N.W. conceived the study. N.W. designed the structure and conducted the experiments while K.C. and N.W. analyzed the experimental results. K.C. and Y.H. supervised the project. K.C. and N.W. wrote the paper. Q.X., X.F., F.L., W.Z. and Y.H. discussed the results and reviewed the manuscript.

**References**

[1] Miri, M.-A. & Alù, A. Exceptional points in optics and photonics. *Science* **363,** eaar7709 (2019).

[2] Özdemir, Ş. K., Rotter, S., Nori, F. & Yang L. Parity–time symmetry and exceptional points in photonics. *Nat. Mater.* **18,** 783–798 (2019).

[3] El-Ganainy et al. Non-Hermitian physics and PT symmetry. *Nat. Phys.* **14,** 11–19 (2018).

[4] Hodaei, H. et al. Enhanced sensitivity at higher-order exceptional points. *Nature* **548,** 187–191 (2017).

[5] Chang, L. et al. Parity–time symmetry and variable optical isolation in active–passive-coupled microresonators. *Nat. Photon* **8,** 524–529 (2014).

[6] Peng, B et al. Parity–time-symmetric whispering-gallery microcavities. *Nat. Phys.* **10,** 394–398 (2014).

[7] Wang, C., Sweeney, W. R., Stone, A. D. & Yang, L. Coherent perfect absorption at an exceptional point. *Science* **373**, 1261–1265 (2021).

[8] Dembowski, C. et al. Experimental observation of the topological structure of exceptional points. *Phys. Rev. Lett.* **86,** 787–790 (2001).

[9] Zhen, B et al. Spawning rings of exceptional points out of Dirac cones. *Nature* **525,** 354–358 (2015).

[10] Park, J.-H. et al. Symmetry-breaking-induced plasmonic exceptional points and nanoscale sensing. *Nat. Phys.* **16,** 462–468 (2020)

[11] Hodaei, H., Miri, M.-A., Heinrich, M., Christodoulides, D. N. & Khajavikhan, M. Parity-time-symmetric microring lasers. *Science* **346,** 975–978 (2014).

12 Lai, Y.-H., Lu, Y.-K., Suh, M.-G., Yuan, Z. & Vahala, K. Observation of the exceptional-point-enhanced Sagnac effect. *Nature* **576,** 65–69 (2019).

13 Hokmabadi, M. P., Schumer, A., Christodoulides, D. N. & Khajavikhan, M. Non-Hermitian ring laser gyroscopes with enhanced Sagnac sensitivity. *Nature* **576,** 70–74 (2019).

14 Chen, W., Kaya Özdemir, Ş., Zhao, G., Wiersig, J. & Yang, L. Exceptional points enhance sensing in an optical microcavity. *Nature* **548,** 192–196 (2017).

15 Xu, H., Mason, D., Jiang, L. & Harris, J. G. E. Topological energy transfer in an optomechanical system with exceptional points. *Nature* **537,** 80–83 (2016).

16 Xu, Z., Gao, X., Bang, J., Jacob, Z. & Li, T. Non-reciprocal energy transfer through the Casimir effect. *Nat. Nanotechnol.* (2021).

17 Doppler, J. et al. Dynamically encircling an exceptional point for asymmetric mode switching. *Nature* **537,** 76–79 (2016).

18 Xu, H., Jiang, L., Clerk, A. A. & Harris, J. G. E. Nonreciprocal control and cooling of phonon modes in an optomechanical system. *Nature* **568**, 65–69 (2019).

19 Barzanjeh, S. et al. Optomechanics for quantum technologies. *Nat. Phys.* **18**, 15–24 (2022).

20 Sansa, M. Sansa et al. Optomechanical mass spectrometry. *Nat. Commun.* **11,** 3781 (2020).

21 Xiong, H., Liu, Z.-X. & Wu, Y. Highly sensitive optical sensor for precision measurement of electrical charges based on optomechanically induced difference-sideband generation. *Opt. Lett.* **42,** 3630 (2017).

22   Kim, P. H., Hauer, B. D., Doolin, C., Souris, F. & Davis, J. P. Approaching the standard quantum limit of mechanical torque sensing. *Nat. Commun.* **7,** 13165 (2016).

23   Lin, Q. et al. Coherent mixing of mechanical excitations in nano-optomechanical structures. *Nature Photon* **4,** 236–242 (2010).

24   Shkarin, A. B. et al. Optically Mediated Hybridization between Two Mechanical Modes. *Phys. Rev. Lett.* **112,** 013602 (2014).

25   Spethmann, N., Kohler, J., Schreppler. S., Buchmann, L. & Stamper-Kurn, D. M. Cavity-mediated coupling of mechanical oscillators limited by quantum back-action. *Nat. Phys.* **12,** 27–31 (2016).

26   Wang, Y.-P. et al. Nonreciprocity and Unidirectional Invisibility in Cavity Magnonics. *Phys. Rev. Lett*. **123**, 127202 (2019).

27   Aspelmeyer, M., Kippenberg, T. J. & Marquardt, F. Cavity optomechanics. *Rev. Mod. Phys.* **86,** 1391–1452 (2014).

28   Kippenberg, T. J. & Vahala, K. J. Cavity Optomechanics: Back-Action at the Mesoscale. *Science* **321,** 1172–1176 (2008).

29   Choi, Y., Hahn, C., Yoon, J. W. & Song, S. H. Observation of an anti-PT-symmetric exceptional point and energy-difference conserving dynamics in electrical circuit resonators. *Nat. Commun.* **9,** 2182 (2018).

30   Li, Y. et al. Anti–parity-time symmetry in diffusive systems. *Science* **364,** 170–173 (2019).

31   Peng, P. et al. Anti-parity–time symmetry with flying atoms. *Nat. Phys.* **12,** 1139–1145 (2016).

32  Chan, J., Safavi-Naeini, A. H., Hill, J. T., Meenehan, S. & Painter. O. Optimized optomechanical crystal cavity with acoustic radiation shield. *Appl. Phys. Lett.* **101,** 081115 (2012).

33  Eichenfield, M., Chan, J., Camacho, R. M., Vahala, K. J. & Painter, O. Optomechanical crystals. *Nature* **462,** 78–82 (2009).

34  Johnson, T. J., Borselli, M. & Painter, O. Self-induced optical modulation of the transmission through a high-Q silicon microdisk resonator. *Opt. Express* **14,** 817 (2006).

35  Cazier, N., Checoury, X., Haret, L.-D. & Boucaud, P. High-frequency self-induced oscillations in a silicon nanocavity. *Opt. Express* **21,** 13626 (2013).

36  Chan, J. et al. Laser cooling of a nanomechanical oscillator into its quantum ground state. *Nature* **478,** 89–92 (2011).

37  Shomroni, I., Qiu, L., Malz, D., Nunnenkamp, A. & Kippenberg, T. J. Optical backaction-evading measurement of a mechanical oscillator. *Nat. Commun.* **10,** 2086 (2019).

38  Safavi-Naeini, A. H. et al. Electromagnetically induced transparency and slow light with optomechanics. *Nature* **472,** 69–73 (2011).

39  Cohen, J. D. et al. Phonon counting and intensity interferometry of a nanomechanical resonator. *Nature* **520**, 522–525 (2015).

40  Burek, M. J et al. Diamond optomechanical crystals. *Optica* **3,** 1404 (2016).

41  Cui, K. et al. Phonon lasing in a hetero optomechanical crystal cavity. *Photon. Res*. **9**, 937 (2021).

42  Zhi, Y., Yu, X.-C., Gong, Q., Yang, L. & Xiao, Y.-F. Single nanoparticle detection using optical microcavities. *Adv. Mater.* **29,** 1604920 (2017).

43  Yu, W., Jiang, W. C., Lin, Q., & Lu, T. Cavity optomechanical spring sensing of single molecules. *Nat Commun* **7,** 12311 (2016).

44  Pan, F. et al. Radiation-Pressure-Antidamping Enhanced Optomechanical Spring Sensing. *ACS Photonics* **5,** 4164–4169 (2018).

45  Massel, F. et al. Multimode circuit optomechanics near the quantum limit. *Nat Commun* **3,** 987 (2012).

46  Ding, K., Ma, G., Xiao, M., Zhang, Z. Q. & Chan, C. T. Emergence, Coalescence, and Topological Properties of Multiple Exceptional Points and Their Experimental Realization. *Phys. Rev. X* **6**, 021007 (2016).

47  Zhang, M. et al. Synchronization of Micromechanical Oscillators Using Light. *Phys. Rev. Lett.* **109,** 233906 (2012).

# Supplementary Information

# On-chip mechanical exceptional points based on an optomechanical zipper cavity:

**I. Theoretical model**

According to the multimode optomechanical model, the Hamiltonian of two mechanical oscillators with resonant frequencies $\Omega_1$, $\Omega_2$ interact with a single optical mode with resonant frequency $\omega_{cav}$ via dispersive coupling can be expressed as[1,2]

$$\hat{H} = \hbar\omega_{cav}\hat{a}^{\dagger}\hat{a} + \sum_{j=1,2}\hbar\Omega_j\hat{b}^{\dagger}_j\hat{b}_j - \sum_{j=1,2}\hbar g_j\hat{a}^{\dagger}\hat{a}(\hat{b}_j+\hat{b}^{\dagger}_j) + i\hbar\sqrt{k_{ex}/2}\alpha_{in}(e^{-i(\omega_L t+\theta)}\hat{a}^{\dagger} - e^{i(\omega_L t+\theta)}\hat{a}) \quad (S1)$$

where $\hbar$ is Planck's constant. $\hat{a}$ and $\hat{b}_j$ represent the annihilation operator of the optical mode and the $j$th mechanical mode, respectively, and $g_j$ is the vacuum optomechanical coupling rate between the $j$th mechanical mode and the optical mode, with $g_j = -(\partial\omega_{cav}/\partial x_j)x_{zpf}(j)$. It characterizes the degree of the frequency shift of the optical resonant mode due to the zero-point fluctuations $x_{zpf}(j)$ of the $j$th mechanical oscillator. The pump laser with frequency $\omega_L$, amplitude $\alpha_{in}$, and phase $\theta$ drives the optical mode through one side of the tapered fiber at a rate of $k_{ex}/2$. Next, we rotate the frame via unitary transformation $\hat{U} = e^{i\omega_L \hat{a}^{\dagger}\hat{a}t}$ to generate a new time-independent Hamiltonian

$$\hat{H} = -\hbar\Delta\hat{a}^{\dagger}\hat{a} + \sum_{j=1,2}\hbar\Omega_j\hat{b}^{\dagger}_j\hat{b}_j - \sum_{j=1,2}\hbar g_j\hat{a}^{\dagger}\hat{a}(\hat{b}_j+\hat{b}^{\dagger}_j) + i\hbar\sqrt{k_{ex}/2}\alpha_{in}(e^{-i\theta}\hat{a}^{\dagger} - e^{i\theta}\hat{a}) \quad (S2)$$

where detuning $\Delta = \omega_L - \omega_{cav}$. In reality, this system is also coupled to both the mechanical and the optical thermal bath. Thus, Hamiltonian $\hat{H}_{bath}$ of the thermal bath and the bath-system interaction term, $\hat{H}_{bath\text{-}system}$, must be included[1,2]. As a result, the quantum Langevin equations that describe the system dynamics can be represented as

$$\frac{d\hat{a}}{dt} = (i\Delta + i\sum_{j=1,2} g_j(\hat{b}_j + \hat{b}^\dagger_j) - k/2)\hat{a} + \sqrt{k_{ex}/2}\alpha_{in}e^{-i\theta} + \sqrt{k}\hat{a}_{in}$$

$$\frac{d\hat{b}_1}{dt} = -(i\Omega_1 + \gamma_1/2)\hat{b}_1 + ig_1\hat{a}^\dagger a + \sqrt{\gamma_1}\hat{b}_{in}(1) \quad\quad (S3)$$

$$\frac{d\hat{b}_2}{dt} = -(i\Omega_2 + \gamma_2/2)\hat{b}_2 + ig_2\hat{a}^\dagger a + \sqrt{\gamma_2}\hat{b}_{in}(2)$$

where the optical vacuum noise $\hat{a}_{in}$ from the thermal bath couple to the cavity at the rate of $k$, while mechanical modes at the thermal states $\hat{b}_{in}(1)$ and $\hat{b}_{in}(2)$ are coupled to the corresponding mechanical modes at the rates of $\gamma_1$ and $\gamma_2$, respectively. In the procedure for solving the dynamics problem in the system, first, we only consider the excitation of the strong pump laser and ignore other excitation terms. The corresponding steady solutions of the system are

$$a_0 = \frac{\sqrt{k_{ex}/2}}{k/2 - i\Delta_{new}}\alpha_{in}e^{-i\theta} \quad \hat{b}_{01} = \frac{ig_1 n_{cav}}{\gamma_1/2 + i\Omega_1} \quad \hat{b}_{02} = \frac{ig_2 n_{cav}}{\gamma_2/2 + i\Omega_2} \quad\quad (S4)$$

where $n_{cav}$ is the average intracavity photon number and $n_{cav} = a_0^* a_0$. Here, we can always change the phase $\theta$ by modifying the reference time to make $a_0$ a real number ($a_0 = \sqrt{n_{cav}}$) and simplify the calculation. We re-define the detuning as $\Delta_{new} = \Delta + \sum_{j=1,2} g_j(\hat{b}_{0j} + \hat{b}_{0j}^\dagger)$ because the radiation pressure of the pump light shifts the cavity resonant frequency.

Regarding the range of the physical parameters used in our experiment, the frequency shift is smaller than several MHz, thus, as influence is insignificant, we can regard $\Delta_{new}$ as $\Delta$. Now, we focus on the dynamics of time-dependent terms. Here we can define $\hat{a} = a_0 + \delta\hat{a}$, $\hat{b}_j = \hat{b}_{0j} + \delta\hat{b}_j$ and insert it into equation (S3) to subtract the steady-state terms. Moreover, as $\delta\hat{a} \ll \hat{a}_0$ in the experiment, we linearize the differential equations by neglecting the high order small terms such as $\delta\hat{a}\delta\hat{b}$, $\delta\hat{a}^\dagger \delta\hat{a}$ [1,2]

$$\frac{d\delta\hat{a}}{dt} = (i\Delta - k/2)\delta\hat{a} + \sum_{j=1,2} ig_j \sqrt{n_{cav}}(\hat{b}_j + \hat{b}^\dagger_j) + \sqrt{k}\hat{a}_{in}$$

$$\frac{d\delta\hat{b}_1}{dt} = -(i\Omega_1 + \gamma_1/2)\delta\hat{b}_1 + ig_1\sqrt{n_{cav}}(\delta\hat{a} + \delta\hat{a}^\dagger) + \sqrt{\gamma_1}\hat{b}_{in}(1) \quad (S5)$$

$$\frac{d\delta\hat{b}_2}{dt} = -(i\Omega_2 + \gamma_2/2)\delta\hat{b}_2 + ig_2\sqrt{n_{cav}}(\delta\hat{a} + \delta\hat{a}^\dagger) + \sqrt{\gamma_2}\hat{b}_{in}(2)$$

Next, the equations (S5) are solved in the frequency domain, where the relation between $\delta\hat{a}$ and $\delta\hat{b}_j$ is

$$\delta\hat{a}(w) = \frac{\sum_{j=1,2} ig_j\sqrt{n_{cav}}(\delta\hat{b}_j(w) + \delta\hat{b}_j^\dagger(w)) + \sqrt{k}\hat{a}_{in}(w)}{k/2 - i(\Delta + w)} \quad (S6)$$

We can substitute the operator $\delta\hat{a}^\dagger$, $\delta\hat{a}$ in the dynamic equations of mechanical oscillators with equation (S6). Therefore, the dynamic equations of mechanical oscillators can be expressed as

$$(H - wI)X = d \quad (S7)$$

where $H=$

$$\begin{bmatrix} \Omega_1 - i\gamma_1/2 + g_1^2\chi & g_1^2\chi & g_1g_2\chi & g_1g_2\chi \\ -g_1^2\chi & -\Omega_1 - i\gamma_1/2 - g_1^2\chi & -g_1g_2\chi & -g_1g_2\chi \\ g_1g_2\chi & g_1g_2\chi & \Omega_2 - i\gamma_2/2 + g_2^2\chi & g_1^2\chi \\ -g_1g_2\chi & -g_1g_2\chi & -g_1^2\chi & -\Omega_2 - i\gamma_1/2 - g_2^2\chi \end{bmatrix} \quad (S8)$$

where $\chi = n_{cav}\left(1/((w+\Delta)+ik/2) - 1/((w-\Delta)+ik/2)\right)$ and $X^T = [\hat{b}_1(w), \hat{b}_1^\dagger(w), \hat{b}_2(w), \hat{b}_2^\dagger(w)]$. $D$ is the excitation term and $d^T \approx -i[\sqrt{\gamma_1}\hat{b}_{in}(1,w), \sqrt{\gamma_1}\hat{b}_{in}^\dagger(1,w), \sqrt{\gamma_2}\hat{b}_{in}(2,w), \sqrt{\gamma_2}\hat{b}_{in}^\dagger(2,w)]$. Here the excitation of the optical vacuum noise is ignored. In our experiment, the coupling between the creation operator $\delta\hat{b}_j$ ($j=1, 2$) and the annihilation operator $\delta\hat{b}_l^\dagger$ ($l=1, 2$) is negligible because the off-diagonal element is much smaller than the difference of the corresponding diagonal element ($g_jg_l\chi \ll (\Omega_j + \Omega_l)$). When $w$ is close to the resonant frequency $\Omega_j$, the evolution of the annihilation operators $\delta\hat{b}_j$ is

$$w\begin{bmatrix} \delta\hat{b}_1(w) \\ \delta\hat{b}_2(w) \end{bmatrix} = \begin{bmatrix} \Omega_1 - i\gamma_1/2 + g_1^2\chi & g_1g_2\chi \\ g_1g_2\chi & \Omega_2 - i\gamma_2/2 + g_2^2\chi \end{bmatrix} \begin{bmatrix} \delta\hat{b}_1(w) \\ \delta\hat{b}_2(w) \end{bmatrix} + i\begin{bmatrix} \sqrt{\gamma_1}\hat{b}_{in}(1,w) \\ \sqrt{\gamma_2}\hat{b}_{in}(2,w) \end{bmatrix} \quad (S9)$$

Here the diagonal elements represent the complex resonant frequencies of the two mechanical oscillators. The complex mechanical frequencies are modified by the optical radiation pressure, similar to single mode optomechanical coupling model. However, the non-zero off-diagonal elements imply that the optical mode also acts as a bridge to connect both mechanical modes in this multimode optomechanical coupling model.

In this two-level system, the eigenvalues represent the dynamics of the mechanical oscillators. The eigenvalues of the system and their relation to the optical parameters are obtained by solving the eigenvalue equations

$$\begin{vmatrix} \Omega_1 - i\gamma_1/2 + g_1^2\chi - \tilde{\omega} & g_1g_2\chi \\ g_1g_2\chi & \Omega_2 - i\gamma_2/2 + g_2^2\chi - \tilde{\omega} \end{vmatrix} = 0 \quad (S10)$$

We assume that the deviation of the eigenvalues to the corresponding intrinsic resonant frequency is smaller than the optical dissipation rate ($|\tilde{\omega}-\Omega_1|\ll k$). Thus, we obtain $\chi(\tilde{\omega})\approx\chi(\Omega_1)\approx\chi(\Omega_2)$. Generally, this assumption is satisfied in the weak coupling regime where $g_j\sqrt{n_{cav}}\ll k$. As a result, the eigenvalue of the two-level system can be expressed as

$$\tilde{\omega}_\pm = \omega_\pm - i\gamma_\pm/2 = \frac{1}{2}(\Omega_1 - i\gamma_1/2 + \Omega_2 - i\gamma_2/2) + \frac{1}{2}(g_2^2 + g_1^2)\chi \pm \frac{1}{2}S \quad (S11)$$

The parameter $S$ determines the frequency splitting and $S=\sqrt{A\varepsilon+B\varepsilon^2}$, where $A = 4ig_1g_2\left((\Omega_1-\Omega_2)-i(\gamma_1/2-\gamma_2/2)\right)$, and $B=(g_1^2-g_2^2)^2+4g_1^2g_2^2$. $\varepsilon$ represents the distance from the parameter $\chi$ to the desired value $\chi_{eps}$ which drives the system into the mechanical EPs ($\varepsilon=\chi-\chi_{eps}$) for the blue-detuned laser used in this work. The EPs are reached when $\chi = \chi_{eps}$ and

$$\chi_{eps} = \frac{((\Omega_1-\Omega_2)-i(\gamma_1/2-\gamma_2/2))}{g_2^2 - g_1^2 - i2g_1g_2} \quad (\Omega_1 > \Omega_2) \quad (S12)$$

Near the EPs, the relation between $S$ and the complex perturbation $\varepsilon$ corresponds to the Riemann surface of $S(\varepsilon)=\sqrt{A\varepsilon}$. Therefore, the amplitude of frequency splitting $|S|$ scales with the 1/2 order of the small perturbation strength $|\varepsilon|$ and this effect has been demonstrated to enhance the sensitivity of sensors by using the optical EPs[3].

Next, we deduce the eigenvalue from the mechanical spectra. According to equation (S6), the optical annihilation operator $\delta\hat{a}$ is related to the mechanical operators $\delta\hat{b}_j$ and $\delta \hat{b}_j^\dagger$. Furthermore, by solving equation (S9), the mechanical annihilation operators $\delta \hat{b}_j$ are obtained as

$$\hat{b}_1(w) = i \frac{-(\Omega_2 - i\gamma_2/2 + g_2^2 \chi - w)\sqrt{\gamma_1}\hat{b}_{in}(1,w) + g_1 g_2 \chi \sqrt{\gamma_2}\hat{b}_{in}(2,w)}{(\tilde{\omega}_+ - w)(\tilde{\omega}_- - w)}$$
$$\hat{b}_2(w) = i \frac{g_1 g_2 \chi \sqrt{\gamma_1}\hat{b}_{in}(1,w) - (\Omega_1 - i\gamma_1/2 + g_1^2 \chi - w)\sqrt{\gamma_2}\hat{b}_{in}(2,w)}{(\tilde{\omega}_+ - w)(\tilde{\omega}_- - w)} \quad (S13)$$

Consequently, on combining equations (S6) and (S13) and ignoring the optical vacuum noise, the optical annihilation operator $\delta \hat{a}$ is a function of the mechanical eigenvalues $\tilde{\omega}_\pm$

$$\delta \hat{a}(w) = \frac{\sqrt{n_{cav}}}{(\Delta + w) + ik/2} \sum_{j=1,2} (T_j(w)\hat{b}_{in}(j,w) + T_j^*(-w)\hat{b}_{in}^\dagger(j,w)) \quad (S14)$$

where $T_1(w) = \left( \frac{i\sqrt{\gamma_1}g_1(\Omega_2 - i\gamma_2/2 - w)}{(\tilde{\omega}_+ - w)(\tilde{\omega}_- - w)} \right)$ and $T_2(w) = \left( \frac{i\sqrt{\gamma_2}g_2(\Omega_1 - i\gamma_1/2 - w)}{(\tilde{\omega}_+ - w)(\tilde{\omega}_- - w)} \right)$.

In addition, the input-output theory connects the output and cavity fields[1,2]

$$\hat{a}_{out}(t) = A_{21}(\Delta)\alpha_{in} - \sqrt{\frac{k_{ex}}{2}}\delta \hat{a}(t) \quad (S15)$$

where $A_{21}(\Delta) = e^{-i\theta}(1 - \frac{k_{ex}/2}{k/2 - i\Delta})$. Therefore, the voltage signal detected by an electric spectrum analyzer (ESA) in the AC coupling mode can be written as

$$V(t) = \eta G \hbar \omega_L (\hat{a}_{out}^\dagger(t)\hat{a}_{out}(t) - |A_{21}(\Delta)|^2 |\alpha_{in}|^2)$$
$$\approx -\eta G \hbar \omega_L \alpha_{in} \sqrt{\frac{k_{ex}}{2}} \left( A_{21}(\Delta)^* \delta \hat{a}(t) + A_{21}(\Delta) \delta \hat{a}^\dagger(t) \right) \quad (S16)$$

where $\eta$ is the transmission efficiency of light in the output fiber and $G$ is the conversion gain of photodetector. Finally, substituting equations (S4) and (S14) into equation (S16) and considering the property of the power spectral density for $\hat{b}_{in}(j,w), \hat{b}_{in}^\dagger(j,w)$ ($j=1, 2$) shown below[4],

$$\langle \hat{b}_{in}^\dagger(i,w)\hat{b}_{in}(j,w')\rangle = n_b\delta(w+w')\delta_{ij}$$
$$\langle \hat{b}_{in}(i,w)\hat{b}_{in}^\dagger(j,w')\rangle = (n_b+1)\delta(w+w')\delta_{ij}$$
$$\langle \hat{b}_{in}(i,w)\hat{b}_{in}(j,w')\rangle = 0 \tag{S17}$$
$$\langle \hat{b}_{in}^\dagger(i,w)\hat{b}_{in}^\dagger(j,w')\rangle = 0$$

the single-sided power spectral density of the voltage signal $V$ in the frequency domain is represented as

$$\bar{S}_{PP}(w) = (S_{PP}(w)+S_{PP}(-w)) = \frac{1}{R_L}\int_{-\infty}^{\infty}\langle V^\dagger(w)V(w')+V^\dagger(-w)V(w')\rangle dw'$$
$$\approx \frac{H(\Delta,k,k_{ex})}{|\tilde{\omega}_+ - w|^2|\tilde{\omega}_- - w|^2}\left(g_1^2\gamma_1\left((\Omega_2-w)^2+(\gamma_2/2)^2\right)+g_2^2\gamma_2\left((\Omega_1-w)^2+(\gamma_1/2)^2\right)\right) \tag{S18}$$

where $\langle\rangle$ denotes ensemble average, $R_L$ is the load resistance of the ESA, and $H(\Delta,k,k_{ex})$ is the transduction function[5]

$$H(\Delta,k,k_{ex}) = \frac{2n_b(k_{ex}\eta GP_{in})^2\Delta^2\left((k-k_{ex}/2)^2+w^2\right)/R_L}{\left(\Delta^2+(k/2)^2\right)^2\left((\Delta+w)^2+(k/2)^2\right)\left((\Delta-w)^2+(k/2)^2\right)} \tag{S19}$$

where $P_{in}$ is the input laser power and $P_{in} = \hbar\omega_L|\alpha_{in}|^2$. $n_b$ is the thermal equilibrium occupancy of phonons at frequency $w$ and $n_b = \frac{1}{e^{\hbar w/k_B T}-1} \approx \frac{k_B T}{\hbar w}$.

## II. The designed optomechanical crystal cavity

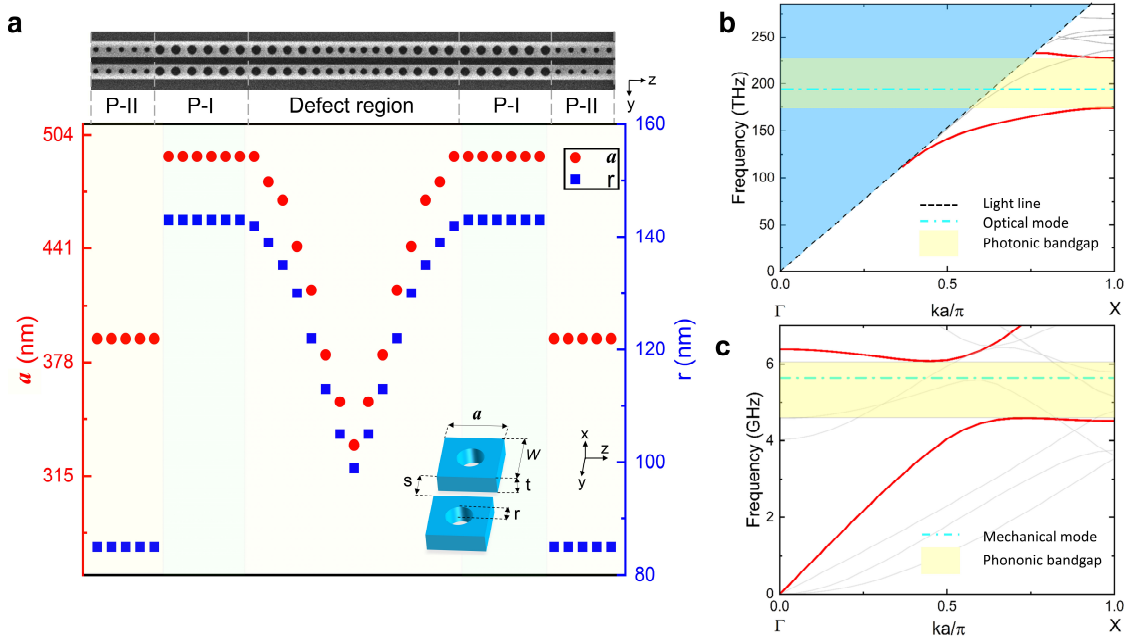

**Figure S1 (a)** Designed structure parameters of unit cells in each region of the optomechanical zipper cavity. **(b)** Optical band structure of unit cell in P-I region. Leaky modes radiate into the environment above the light line (black dash-dotted). Instead, guide modes are confined in dielectric under the light line. Yellow region represents optical bandgap for x-even and y-odd vector symmetry guide modes (red lines), and grey lines show guide modes with other symmetry; Optical resonant frequency (cyan dash-dotted line) is at center of this bandgap. **(c)** Mechanical band structure of the unit cell in P-II region. Yellow region represents mechanical bandgap for x and y symmetry modes (red lines) and mechanical resonant frequency (cyan dash-dotted line) lies in this region. Gray lines show guide modes with other symmetry. Band structures in (b)-(c) are simulated via FEM.

To realize mechanical EPs, a strong optomechanical coupling rate $g_0$ and low optical dissipation coupling rate $k$ are required. Thus, the structure parameters should be designed carefully. In this manuscript, the optomechanical crystal is composed of unit cells with different radius $r$ and lattice constant $a$. The designed radius $r$ and lattice constant $a$ for the designed cavity are shown in Fig. S1(a), with the width of the nanobeam $W$=472 nm, the thickness of the device layer $t$=220 nm, the gap length $s$=200 nm.

To further verify the effectiveness of the photonic and phononic periodic "mirror," we concentrate on the band structure in one arm of the optomechanical zipper cavity for simplicity. The optical band structure of the unit cell of one arm in the P-I region is shown in Fig. S1(b), wherein the resonant frequency of the x-even and y-odd vector symmetry optical mode is localized at the center of the quasi-bandgap of the guide modes. This indicates that the optical mode is confined without dissipating through the waveguide. The mechanical band structure of the unit cell of one arm in the P-II region is demonstrated in Fig. S1(c), and the mechanical resonant mode with x- and y-symmetry is confined by the bandgap of guide modes with the same symmetry. Furthermore, the control of the mechanical loss can be achieved by changing the radius of the holes in P-II region to shift the frequency range of the bandgap in Fig. S1(c).

## III. Fabrication process of zipper cavity

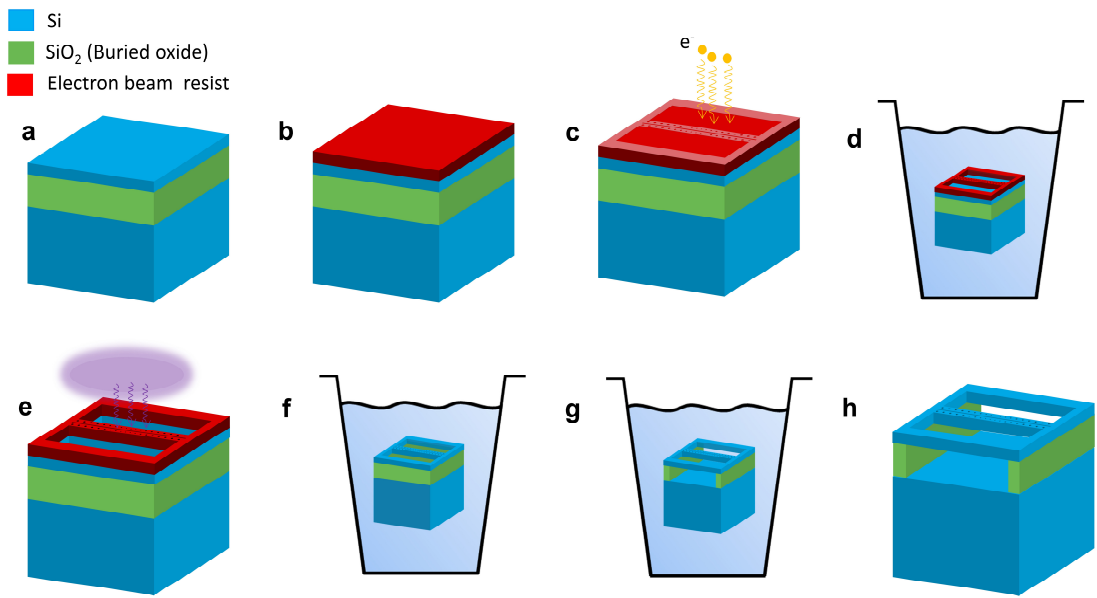

**Figure S2 Fabrication process of devices (a)** A 220 nm thick silicon layer sets on top of 3 μm buried oxide layer. **(b)** Spin electron beam resist. **(c)** Define pattern on the resist by electron beam lithography (EBL). **(d)** Develop and rinse the resist. **(e)** Etching silicon layer and transferring the pattern via inductively coupled plasma etching (ICP). **(f)** Remove the electron beam resist. **(g)** Remove the buried oxide layer using buffered hydrofluoric acid. **(h)** Final designed structure.

The fabrication process of the Zipper cavity is shown in Fig. S2. After determining the parameters of the zipper cavity, the designed cavity was fabricated on a silicon-on-insulator wafer having top layer thickness of 220 nm and a 3-μm thick buried-oxide layer. Electron-beam lithography (EBL) was used to define the pattern on resist and inductively coupled plasma reactive ion etch (ICP) was employed to transfer the pattern to the top silicon layer. Subsequently, the buried-oxide layer was removed using buffered hydrofluoric acid (BHF) to form the suspended nanobeam structure.

## IV. Experimental setup

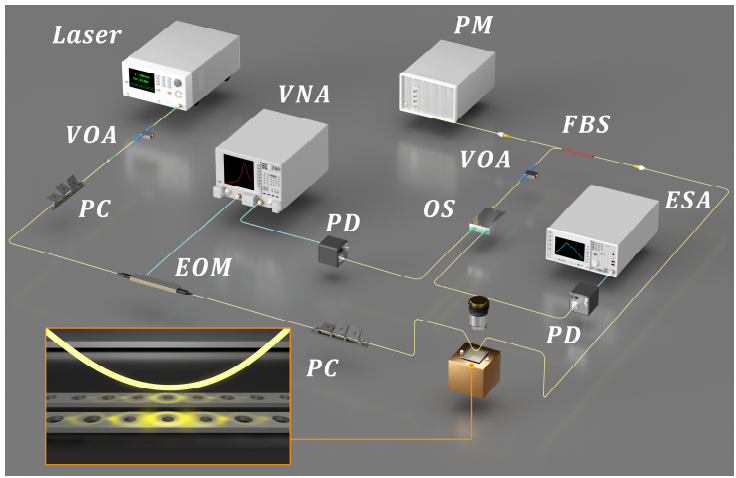

**Figure S3 Experimental setup schematic.** VOA: variable optical attenuator; PC: polarization controller; EOM: electro-optic modulator; VNA: vector network analyzer; FBS: fiber beam splitter; PM: power meter; OS: optical switch; PD: photodetector; ESA: electric spectrum analyzer.

The measurement setup used to characterize the evolution of mechanical spectra with tuned optical parameters is shown in Fig. S3. A tapered fiber controlled with nanopositioners was used to evanescently couple the light into the zipper cavity. The pump light originated from a tunable diode laser, and the light power was adjusted by a variable optical attenuator (VOA). Thereafter, two polarization controllers (PC) were adjusted to excite an optical resonant mode efficiently through the extrinsic coupling channel at rate $k_{ex}/2$. Since the motion of mechanical oscillators modulated the pump laser in the cavity, resulting in the generation of scattering light that contained mechanical information, the beating between the pump light and the scattering light was collected via a photodetector (PD). Subsequently, it was analyzed using an electric spectrum analyzer (ESA). Moreover, the transmitted power of the pump light was monitored by a power meter (PM). In this experiment, a pump-probe scheme was used to monitor the detuning $\Delta$ and the optical dissipation rate $k$. For this scheme, the pump light was modulated by the phase electro-optic modulator (EOM) driven by a vector network analyzer (VNA). The generated probe light passes through the optomechanical cavity and was collected by another PD. Here the VOA in front of the optical switch (OS) was used to avoid the saturation of the PD. Finally, the VNA demodulated the electric signals to obtain response curves of the optical cavity. The real-time monitoring via VNA is essential because the optical resonant frequency $\omega_{cav}$ and the dissipation rate $k$ rely on the intracavity photon number $n_{cav}$ owing to the thermal-optic effect and the nonlinear absorption processes[6], respectively.

## V. Fitting frequency response in pump-probe experiment

In the pump-probe experiment, the pump laser is modulated by driving the phase modulator with RF signal from the VNA. One of the generated sidebands scans across the optical cavity. Consequently, the beating between the pump-laser and the sidebands is detected by the photodetector and analyzed by the VNA, as shown in Fig. S3. When ignoring the optomechanical coupling, the magnitude of the $S_{21}$ response can be expressed as[7]

$$|S_{21}(f)|=\beta(f)\left|\left(A_{21}(\Delta)A_{21}^*(\Delta-2\pi f)-A_{21}^*(\Delta)A_{21}(\Delta+2\pi f)\right)\right| \qquad (S20)$$

where $f$ is the modulation frequency. Here, both the modulation coefficient of the phase modulator and the gain of the photodetector are frequency-dependent, and they are considered in the frequency-dependent coefficient $\beta(f)$.

To fit the detuning $\Delta$ and the optical dissipation rate $k$ precisely from the magnitude of the $S_{21}$ response, the coefficient $\beta(f)$ is required to calibrate the $S_{12}$ response. In our experiment ($\Delta \gg k \gg k_{ex}$), if one sideband is close to the optical resonance ($|\Delta-2\pi f| \ll k$), then equation (S20) can be simplified as

$$|S_{21}(f)|=\beta(f)\left|(k_{ex}/2)/(k/2-i(\Delta-2\pi f))\right| \qquad (S21)$$

The maximum possible value of the spectrum $S_{21}(f)$ is $S_{21}(f=\Delta/2\pi)=(k_{ex}/k)\beta(f)$. Therefore, the peak value of $S_{21}$ at different detuning $\Delta$ scales with $\beta(\Delta/2\pi)$, and the normalized $\beta(f)$ can be obtained. As shown in Fig. S4, the normalized coefficient $\beta(f)$ is derived from the experimental $S_{21}$ spectra at low pump power where the nonlinear absorption processes are negligible, and the optical dissipative rate $k$ is a constant. Subsequently, the $S_{21}$ spectra at high pump power are calibrated and fitted using equation (S20) to acquire $k$ and $\Delta$ at real-time.

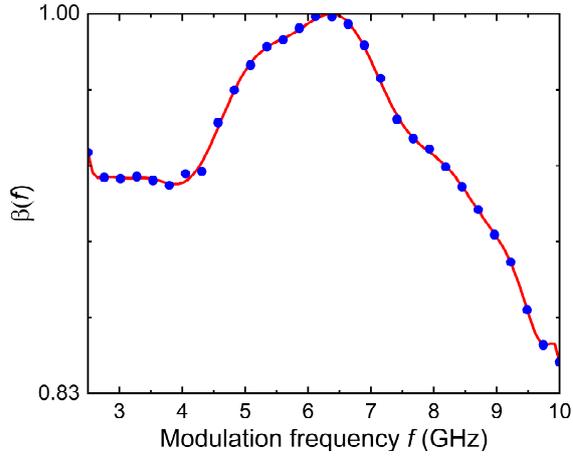

**Figure S4.** Frequency-dependent coefficient $\beta$ versus modulation frequency $f$. The blue dots represent the experimental data and red curve corresponds to the polynomial fitting.

## VI. Influence of nonlinear absorption effect.

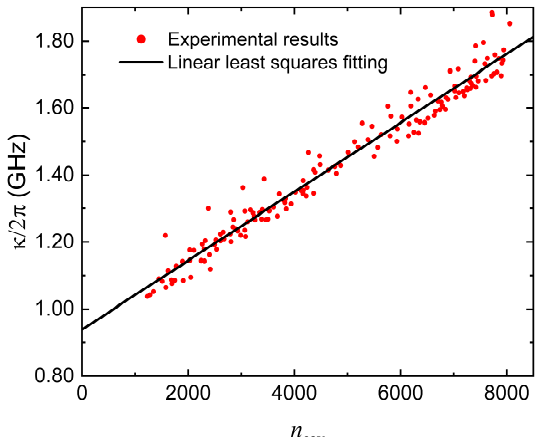

**Figure S5.** Optical dissipation rate $k$ at different intracavity photon numbers $n_{cav}$. Red dots are speculated from the obtained experimental spectra of $S_{21}$, and they are fitted using the linear least square method shown in black line.

The optical dissipation rate $k$ in the silicon zipper cavity depends on the intracavity photon number $n_{cav}$ owing to the two-photon absorption (TPA) and the free-carrier absorption (FCA) effects[6]. As shown in Fig. S5, the relation between $k$ and $n_{cav}$ indicates that $k$ increases linearly with $n_{cav}$. The fitting results of the nonlinear absorption effect are considered when calculating the theoretical results in Figs. 3c-f and 4a-b of the main text.

## VII. Mechanical frequency drift

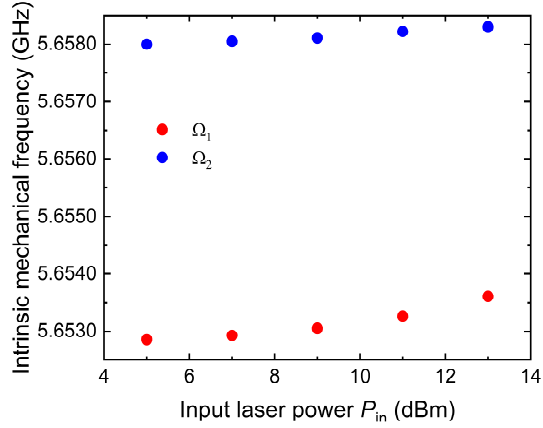

**Figure S6.** The fitting intrinsic mechanical frequencies under different input light power during experiment.

In our experiment, we varied the light power individually and scanned the detuning Δ for each light power to obtain the mechanical spectra. During the measurement, mechanical frequency drift occurred. The intrinsic mechanical frequencies were calibrated under each input power and are shown in Fig. S6. This mechanical frequency drift is considered in the theoretical line in Figs. 3c-f of the main text. We also observed the mechanical frequency drift when characterizing other single nanobeam optomechanical crystals independently. This long-term drift may be attributed to the variation of Young's modulus for unknown reasons, which needs to be investigated further.

## VIII. Experimental results in the phonon lasing regime

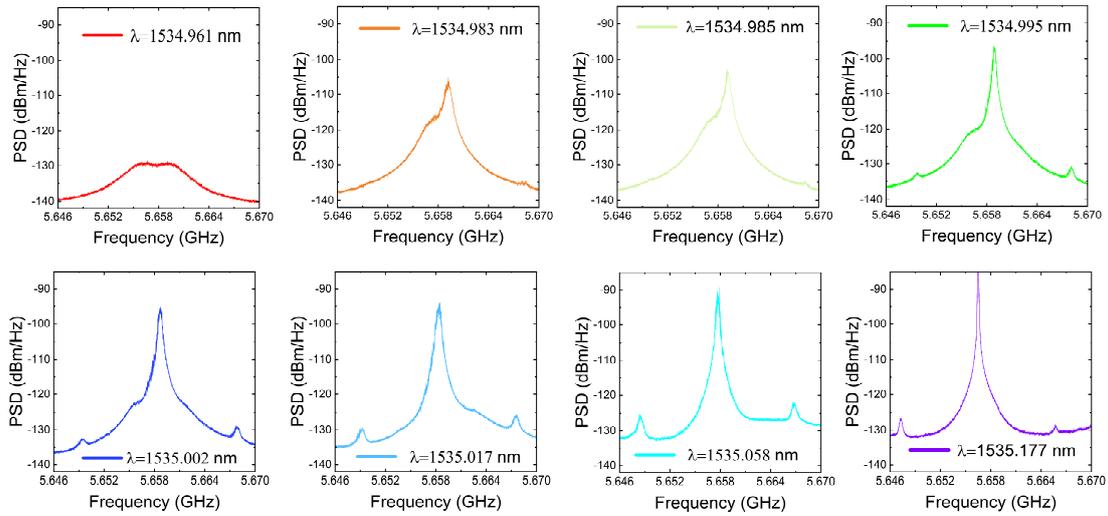

**Figure S7.** The evolution of the mechanical spectra shape from the linear regime to the phonon lasing regime with high laser power.

In the main text, only the experimental results before entering the phonon lasing regime are demonstrated. Figure S7 displays the power spectrum density (PSD) of mechanical spectra in an independent test in this sample with high laser power. As mentioned in the main text, the linewidth of mechanical mode with higher frequency decreases dramatically when increase (decrease) the wavelength (detuning). Subsequently, the phonon lasing of the high-frequency mode happened. It is noted that the mechanical spectra in the phonon lasing regime are beyond the quantitative description in equation (S18). When we increase the laser wavelength further, the peak of the mechanical modes with lower frequency will decrease, and the contrast between these two modes will increase. Meanwhile, the difference in their mechanical frequencies also changes. Finally, only one peak was observed when further increases the wavelength.

**References**

1   Aspelmeyer, M., Kippenberg, T. J. & Marquardt, F. Cavity optomechanics. *Rev. Mod. Phys.* **86,** 1391–1452 (2014).

2   Bowen, W. P. Bowen & Milburn, G. J. *Quantum Optomechanics* (CRC Press, Taylor & Francis Group, Boca Raton, 2016).

3   Özdemir, Ş. K., Rotter, S., Bori, F. Nori & Yang, L. Parity–time symmetry and exceptional points in photonics. *Nat. Mater.* **18,** 783–798 (2019).

4   Chan, J. Chan et al. Laser cooling of a nanomechanical oscillator into its quantum ground state. *Nature* **478,** 89–92 (2011).

5   Gorodetsky, M. L., Schliesser, A., Anetsberger, G., Deleglise, S. & Kippenberg, T. J. Determination of the vacuum optomechanical coupling rate using frequency noise calibration. *Opt. Express* **18,** 23236 (2010).

6   Johnson, T. J., Borselli, M. & Painter, O. Self-induced optical modulation of the transmission through a high-Q silicon microdisk resonator. *Opt. Express* **14,** 817 (2006).

7   Shomroni, I., Qiu, L., Malz, D., Nunnenkamp, A. & Kippenberg, T. J. Optical backaction-evading measurement of a mechanical oscillator. *Nat. Commun.* **10,** 2086 (2019).